\documentclass[
twocolumn,
aps,nofootinbib,showpacs,showkeys,preprint
tightenlines,preprintnumbers,
,superscriptaddress
] {revtex4}

\usepackage{epsf,epsfig,subfigure,graphicx,amsmath,amssymb}
\usepackage{color}
\usepackage{float}

\newcommand{\dis}[1]{\begin{equation}\begin{split}#1\end{split}\end{equation}}

 \newcommand{\dell}{\delta_{\rm PMNS}}
\newcommand{\delq}{\delta_{\rm CKM}}

\newcommand{\delx}{\delta_{\rm X}}
\newcommand{\eL}{\epsilon_{\rm L}}

\newcommand\lsim{\mathrel{\rlap{\lower4pt\hbox{\hskip1pt$\sim$}}
    \raise1pt\hbox{$<$}}}
\newcommand\gsim{\mathrel{\rlap{\lower4pt\hbox{\hskip1pt$\sim$}}
    \raise1pt\hbox{$>$}}}

\newcommand\etal{{\it et al.}}

\newcommand\ie{{\it i.e.}~}

\newcommand{\Z}[1]{{\bf Z}}

\begin{document}

\title{Type-II Leptogenesis\footnote{Talk presented at ICHEP 2016, Chicago, U.S.A., 3-10 August 2016.}}

\author{Jihn E. Kim }
\address{ 
Department of Physics, Kyung Hee University, 26 Gyungheedaero, Dongdaemun-Gu, Seoul 02447, Republic of Korea, and \\
 Center for Axion and Precision Physics Research (CAPP, IBS),
  291 Daehakro, Yuseong-Gu, Daejeon 34141, Republic of Korea  
}
   
\begin{abstract}
I discuss on our new theory on baryogenesis  `Type-II leptogenesis' \cite{CoviKim16} which is different from the well-known `Type-I leptogenesis' \cite{FY86}. First, I will briefly comment on the Jarlskog phases  in the CKM and PMNS matrices, $\delq$ and $\dell$. Then, the PMNS phase is used in the `Type-II' leptogenesis for Sakharov's condition on the global quantum number generation in the Universe. For this to be effective, the SU(2)$\times$U(1) gauge symmetry must be broken during the leptogenesis epoch. 
\end{abstract}
\pacs{11.30.Er, 11.30.Hv, 11.30.Fs, 11.10.Wx}

\keywords{PMNS matrix, Leptogenesis, Potential at high temperature}

 \maketitle

\section{Introduction}\label{sect:intro}

\begin{figure}[!b]
  \begin{center}
  \begin{tabular}{c}
   \includegraphics[width=0.45\textwidth]{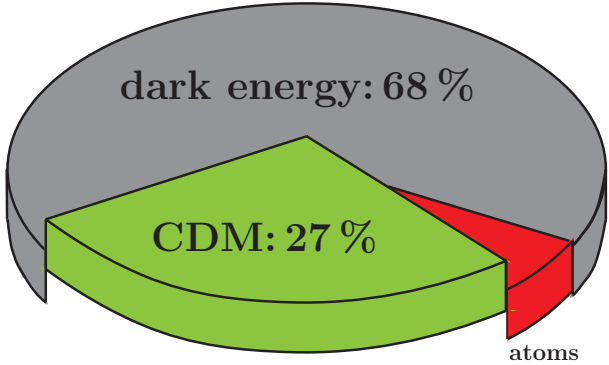}
   \end{tabular}
  \end{center}
 \caption{The energy pie of the Universe.
  }
\label{fig:Epie}
\end{figure}

\begin{figure}[!t]
  \begin{center}
  \begin{tabular}{c}
   \includegraphics[width=0.45\textwidth]{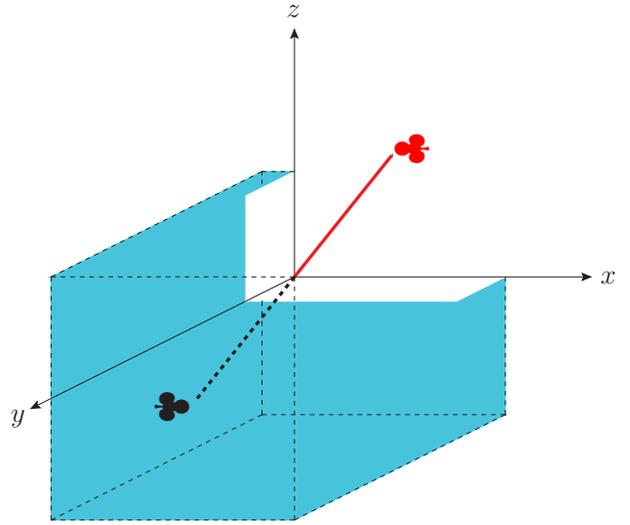}
   \end{tabular}
  \end{center}
 \caption{The ${B}$ number in the parity operated part in the Universe.
  }
\label{fig:Parity}
\end{figure}

In this talk, I discuss the mere 5\,\% of the energy pie (mainly of atoms composed of baryons) of the Universe shown in Fig. \ref{fig:Epie}.  At the most fundamental level of  the grand unification(GUT) scheme, it belongs to the problem on the chiral representation of quarks \cite{Georgi79,KimICHEP1,KimJHEP15}. In cosmology, it belongs to Sakharov's three conditions. The first one is the existence of baryon number ($B$) violating interaction, and the second is the existence of CP violation \cite{Sakharov67}. For example, at the red clover point of Fig. \ref{fig:Parity} let us calculate baryons moving forward/backward in time. The ratio is the number of baryons over the  number of antibaryons. Therefore, to have a nonzero $\Delta B$, we need T or CP violation. We also need to break C. To see this, firstly assume T violation and calculate $\Delta B$ at the red clover point in Fig. \ref{fig:Parity}. Second, integrate over solid angles. At the black clover point, if C is not broken then 
$\Delta B$ is minus that of red clover point, because $P=-1$ is multiplied.  Then, integration over the solid angle gives $\Delta B=0$. Therefore, the needed T violation in the $\Delta B\ne 0$ processes must accompany the C violation also. The CP violation in  the standard model(SM) arises as the ``V-A'' form and hence also breaks C.  Sakharov's third condition on ``out of thermal equilibrium'' is satisfied by the decay of some heavy particle(s), which is usually assumed in most baryo- and lepto-genesis mechanisms.
 
The CP violation seems to work also in some models of dark matter in the Universe. The asymmetric dark matter scenario for this case follows the paradigm of baryogenesis and hence CP violation is the key here also. The axion scenario for dark matter \cite{Baer15} uses an oscillating axion field whose amplitude at present is of order $|\bar{\theta}_{\rm max}|\simeq 10^{-20}$ \cite{KimJeju16}. For nonzero $\bar{\theta}$ though oscillating, CP is violated  and hence CP violation is the key in the axionic dark matter scenario also. Only, the WIMP scenario for dark matter uses just the  cosmological freezeout temperature, not employing any kind of CP violation \cite{Baer15}.

In this talk I will briefly review the weak CP violation first and then present a new mechanism on the leptogenesis that we call `Type-II leptogenesis' with CP violation in the chiral theory \cite{CoviKim16}. 

\section{CP violation and CKM matrix}

In the SM, CP violation is represented by $W_\mu^\pm$ couplings to SU(2)$_W$ doublet fermions which are parametrized by the unitary CKM and PMNS matrices \cite{KM73,PMNS}. In particular, the elements of the CKM matrix are well-known by now, and there are three classes of parametrizations, which we explain below. 
Let the CP phase of the CKM matrix be $\delq$. Then, any Jarlskog triangle \cite{Jarlskog85} has an angle $\delq$ \cite{KimICHEP1}.  The Jarlskog invariant $J$  has a simple expression with Det.\,$V_{\rm CKM}=1$ \cite{KimMoNam15},
 \dis{
 J=|\textrm{Im\,}V_{31}V_{22}V_{13}|.
 }

\begin{figure}[!t]
  \begin{center}
  \begin{tabular}{c}
   \includegraphics[width=0.35\textwidth]{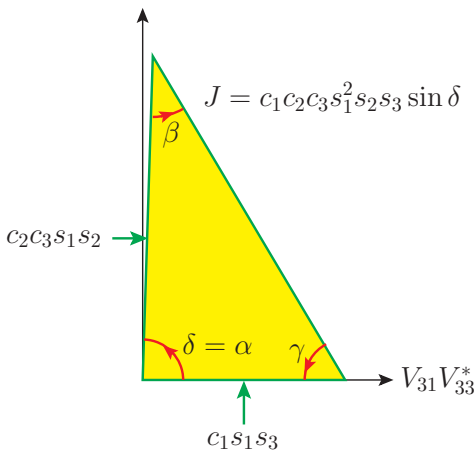}
   \end{tabular}
  \end{center}
 \caption{The Jarlskog triangle of area O($\lambda^6$) from $B$ meson decays with three comparable sizes of O($\lambda^3$)  in the KS parametrization \cite{KimSeo11}.
  }
\label{fig:JTriBmeson}
\end{figure}

\begin{figure}[!t]
  \begin{center}
  \begin{tabular}{c}
   \includegraphics[width=0.30\textwidth]{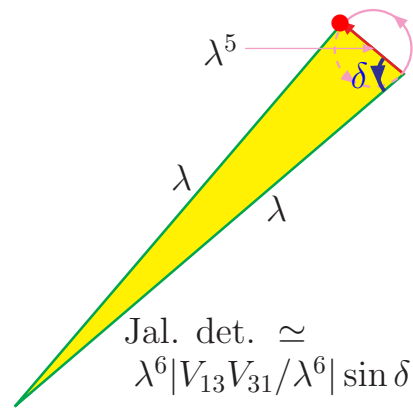}
   \end{tabular}
  \end{center}
 \caption{The Jarlskog triangle of  area O($\lambda^6$) with two long sides of O($\lambda$). This area must be exactly the same as that of Fig. \ref{fig:JTriBmeson}. The angle $\delta$ is rotated and $J$ is maximal with $\delta\simeq 90^{\rm o}$.
  }
\label{fig:JTriMax}
\end{figure}

There are six Jarlskog triangles. A triangle with the lengths of O($\lambda^3$) for all three sides, where $\lambda=\sin\theta_C$, is shown in Fig. \ref{fig:JTriBmeson}. Here we denoted the sides in the Kim-Seo (KS) parametrization  \cite{KimSeo11}.
In this figure, there are three angles, $\alpha,\beta$ and $\gamma$, any of which is not small. We can use any of these as $\delq$, which can be called, `$\alpha$-class', `$\beta$-class', and `$\gamma$-class', respectively. The KS parametrization and the Kobayashi-Maskawa parametrization \cite{KM73} belong to the $\alpha$-class, and the Chau-Keung-Maiani (CKM) parametrization \cite{CK84,Maiani76,PDG15}  belongs to the $\gamma$-class   \cite{KimICHEP1}. 
Note that the unitarity triangle in the PDG book gives \cite{PDG15}
\dis{
&\alpha=\left(85.4^{+3.9}_{-3.8}\right)^{\rm o},
~\beta=\left(21.50^{+0.76}_{-0.74}\right)^{\rm o}\\
&\gamma=\left( 68.0^{+8.0}_{-8.5}\right)^{\rm o}.\label{eq:PDGfit}
 }
 Since $\alpha$ is close to $90^{\rm o}$, we use the KS parametrization,
\dis{
&\left(
\begin{array}{ccc}
c_1 & s_1c_3 & s_1s_3  \\
-c_2s_1 & -e^{-i\delta'}s_2s_3+c_1c_2c_3 & e^{-i\delta'}s_2c_3+c_1c_2s_3  \\
e^{i\delta'}s_1s_2 & -c_2s_3-c_1s_2c_3 e^{i\delta'} & c_2c_3-c_1s_2s_3e^{i\delta'} \\
\end{array}\right), 
 } 
where $\delta'=\delq$, and   $c_i=\cos\theta_i$ and $s_i=\sin\theta_i$. With this parametrization, we can prove that the weak CP violation is maximal with the currently determined real CKM angles. Since any Jarlskog triangle gives the same $J$, consider the triangle with two long sides of O($\lambda$) as shown in Fig. \ref{fig:JTriMax}.
The area of the triangle ($=\frac12 J$) is O($\lambda^6$). So, it is $\frac12 \lambda\cdot\lambda^5\cdot\sin\delta$ which is maximal with $\delta=90^{\rm o}$. With the measured real angles $\theta_{1,2,3}$, this maximality must be true in any parametrization. Even we use the CKM parametrization \cite{CK84}, it must be so but the proof may not be so 
simple\footnote{To prove the maximality with the observed real angles in the CKM parametrization \cite{CK84}, one must slightly vary the real angles also since $\gamma$ is not close to $ \frac{\pi}{2}$.} as given in the KS parametrization where $\delq=\alpha$ is close to $\frac{\pi}{2}$.  This is based on the fact that in a given parametrization  all the six triangles give the identical area. 

Note, however, that at present a specific parametrization is not better than others. If one finds a process to determine three real angles or $\delq$ itself, the measured value(s) for that process can choose a correct class of parametrization scheme.

\section{Maximal CP violation in the PMNS matrix?}

In the previous section, we observed that there is a possibility that $\delq=\frac{\pi}{2}$. It will be very interesting if the leptonic CP violation in the SM also gives  $\dell=\pm  \frac{\pi}{2}$. Then, one may look for a theory relating $\delq$ and $\dell$ as tried in  \cite{KimNam15}. Even though the significance is very low, there is a hint that $\dell\approx -\frac{\pi}{2}$ \cite{T2KCP}. But $\dell\ne 0$ is a more significant statement than fixing it as $\dell\approx -\frac{\pi}{2}$.

The PMNS lepton mixing angles in the SM is parametrized by
\dis{
 &\left(
\begin{array}{ccc}
C_1 & S_1C_3 & S_1S_3  \\
-C_2S_1 & -e^{-i\delta}S_2S_3+C_1C_2C_3 & e^{-i\delta}S_2C_3+C_1C_2S_3  \\
e^{i\delta}S_1S_2 & -C_2S_3-C_1S_2C_3 e^{i\delta} & C_2C_3-C_1S_2S_3e^{i\delta} \\
\end{array}\right)\label{eq:PMNS}
 } 
where  $\delta=\dell$ \cite{PMNS}, and   $C_i=\cos\Theta_i$ and $S_i=\sin\Theta_i$.
Because $\dell$ is non-zero, let us look for leptogenesis mechanism which includes $\dell$ explicitly.
 
\section{Leptogenesis}
It is known that the sphaleron processes at the electroweak scale changes
$B$ and $L$ numbers but conserves $B-L$  \cite{EWgen}. At the electroweak scale, the sphaleron conversion of $L$ to $B$ or vice versa gives the final $B$ and $L$ proportional to the initial $B-L$. Since the SU(5)  GUT  conserves $B-L$, thus the early GUT idea \cite{BAgut} on baryogenesis seems not working.

 For non-zero final baryon number, generation  of  $B$ and $L$ at high temperature must occur through processes which generate nonzero $B-L$. In GUTs, therefore, we use the $B-L$ breaking interaction in
SO(10), for example. In the beyond SM (BSM) with singlets, we can use a singlet $N$, called ``heavy neutrino'' at high energy scale, to generate just $L$. The early idea was to define a right(left)-handed $N$ with lepton number $L=+1(-1)$ \cite{FY86}. In fact, the definition of lepton number, related to neutrino masses, is a combination of defining the lepton number  of the up-type Higgs doublets together with that of $N$. The leptogenesis with the definition of $L=+1$ for $N_R$ is called here `Type-I leptogenesis' \cite{FY86}.

\section{Type-II leptogenesis}

`Type-II leptogenesis' is proposed \cite{CoviKim16} under a different definition on $L$ from that of Type-I and the electroweak symmetry breaking at high temperature. There exists a finite region of parameter space in the multi-Higgs model that the electroweak symmetry is broken at high temperature  \cite{Senj79}. In SUSY, the electroweak symmetry breaking at high temperature is more probable since the temperature dependent terms in $V$ are SUSY breaking terms.

The lepton number is defined with the left-handed SM doublets $\ell_L(\ni\nu_{e,\mu,\tau})$ carrying $L=+1$. Weinberg's effective operator for neutrino masses is \cite{WeinbergNu79}
  \dis{
{\cal L}_Y\propto\frac{f_{ij}}{M}\ell_i\ell_j h_uh_u \label{eq:Wein}
  }
where $h_u$ is the up-type Higgs doublet. Non-zero neutrino masses break $L$. So, as far as $\langle h_u\rangle=0$, neutrinos do not obtain mass by the above operator. Therefore, $L$ can be properly defined below the scale of  $\langle h_u\rangle\ne 0$. One may argue that there is also $h_d$, which is exactly the reason that $h_u$ can carry a global quantum number, because both $h_u$ and $h_d$ can together define a gauge charge $Y$ and a global charge $L$.   To do so, in fact we need `inert' SU(2) doublets $H_{u,d}$.

  The effective interaction (\ref{eq:Wein}) is realized in terms of two renormalizable operators
 \dis{
 M_0NN,~\ell N h_u 
 }
where $M_0$ in our jargon is the `messenger' mass, connecting $N$ and $\nu$,  which is the so-called ``seesaw'' mechanism, first suggested with a humdrum title in \cite{Minkowski77}. But, for neutrino masses at low energy, ``who cares about renormalizable interactions\,?'', as noticed in many mechanisms for neutrino masses of Type$_{I,II,III}$.

\begin{table} 
\centerline{\begin{tabular}{|l| ccc|} 
\hline  &&&\\[-1em]
  &$\ell_L$ &$h_u\,(H_u)$ & $N$\\ \hline\\[-1.4em]
&&&  \\[-1em]
Type-I leptogenesis  &  ~$+1$~ &~$0$~&~$-1$~\\[0.3em] 
Type-II leptogenesis &  ~$+1$~ &~$0\,(-2)$~ &$-1$~\\[0.3em]
\hline
\end{tabular}}
\caption{Definition of lepton numbers, in Type-I and Type-II leptogeneses.}
\label{tab:DefQno}
\end{table}

 However, in cosmology it matters because in the cosmic history the cosmic temperature $T$ has swept all possible energy scales of particle physics. One can define the {\it lepton number} $L$ as shown in Table \ref{tab:DefQno}.
 
\begin{figure}[t!]
  \begin{center}
   \includegraphics[width=0.45\textwidth]{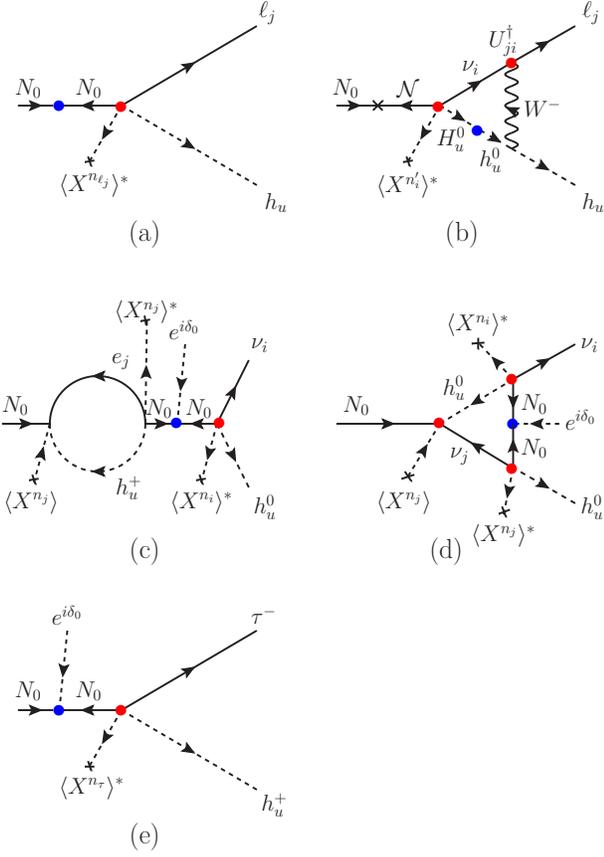}
  \end{center}
 \caption{The Feynman diagrams interfering in the $N$ decay: (a) the lowest order diagram, (b) the $W$ exchange diagram, (c) the wave function renormalization diagram, and (d)  the heavy neutral lepton exchange diagram.  There exist similar ${\cal N}$-decay diagrams.   In all figures, the final leptons can be both charged leptons and neutrinos. 
 The lepton number violations are inserted with blue bullets and phases are inserted at red bullets. (a) and (b) interfere.  (c) and (d) give a vanishing contribution in $N_0$ domination with one complex VEV.} \label{fig:IntCP}
\end{figure}

In Type-I leptogenesis, Figs. \ref{tab:DefQno}\,(e), (c), and (d) are considered where both the $L$ violation and CP violation appear in blue bullets. We need at least two $N$'s for the interference of  Figs. \ref{tab:DefQno}\,(e), (c), and (d). So, if there is a hierarchy of masses such as $m_{N_0}/m_{N_1}\ll 1$, then the lepton asymmetry is suppressed by that factor.
   
\begin{table} 
\centerline{\begin{tabular}{|c| c ccccc|} 
\hline  &&&&&&\\[-1em]
  &~$\ell_L$~ &$H_u$&$H_d$ & $h_{u,d}$&$N$ & ${\cal N}$\\ \hline\\[-1.4em]
&&&&&&  \\[-1em]
 Type-II leptogen. &  ~$+1$~ &~$-2$~&~$+2$~ &~$0$~&~$-1$~&~$+1$~\\[0.3em]
 VEV &  ~$\times$~ &~\textrm{inert}~ & $ \textrm{inert}$  & ~$v_{\rm ew}\{s_\beta,c_\beta\}$ &~$\times$~&~$\times$~\\[0.3em]
\hline
\end{tabular}}
\caption{Definition of lepton numbers  in Type-II leptogeneses. We introduced an inert Higgs $H_u$ carrying $L=-2$ with zero VEV and singlet leptons ${\cal N}$ carrying $L=+1$.}
\label{tab:Type2}
\end{table}

In Type-II leptogenesis, the lepton number is defined differently as shown in Table \ref{tab:Type2}. We need an inert Higgs doublet $H_u$ carrying $L=-2$  and singlet leptons ${\cal N}$ carrying $L=+1$. The relevant leptogenesis diagrams are those given in  Figs. \ref{tab:DefQno}\,(a) and (b). Here, the lepton number violation  appears in blue bullets and the CP violation appears in red bullets. Anyway, the fields $H_{u,d}$ and ${\cal N}$ introduced at high energy scale are not visible at low energy scale. We  introduce interactions
\dis{
&N_0\,\ell_L h_u,~{\cal N}_0\,\ell_L H_u,~N_0\,{\cal N}_0,~H_uH_d,\cdots\\
&h_u^*H_u,~N_0\,N_0,~{\cal N}_0{\cal N}_0, \cdots 
}
where the first line conserves $L$ and  the second line violates $L$.

For Fig. \ref{tab:DefQno}\,(b) to be useful for leptogenesis at a high temperature, we must have $W$ boson is not massless, \ie SU(2)$\times$U(1) symmetry is broken at the high temperature. Indeed, such possibility has been suggested long time ago \cite{Senj79}.
Furthermore, in models with only one CP phase in an ultraviolet completed theory, leptogenesis phase can be related to the
PMNS phase.  Even, phases of heavy neutrinos, $N_0,\cdots$, are expressible in terms of this fundamental phase \cite{KimNam15}.

Within this framework, we calculated the lepton asymmetry which turned out to be
consistent with the fact that in these diagrams 
the lepton violation is on the left side of the cut diagram as discussed in \cite{Adhikari:2001yr}. 
The lepton asymmetry we obtain is a form,
\dis{
\eL^{N_0}(W)&
\approx \frac{\alpha_{\rm em}}{2\sqrt{2}\sin^2\theta_W}  \frac{\Delta m_h^2 }{m_0^2}\\
&\cdot\sum_{i,j}{\cal A}_{ij}\sin[(\pm n_P+n'-n_i+n_j)\delx],
\label{eq:asymmExp4}
}
where ${\cal A}_{ij}$ are given by Yukawa couplings, $n_P,n',n_i,n_j$ are integers, and we assumed that only one phase $\delta_X$ appears in the full theory.  Two independent $n'$ are Majorana phases multiplied to the PMNS matrix.
Using the sphaleron calculation of  \cite{D'Onofrio:2014kta}, we obtain
\dis{
\frac{\Gamma_{\rm sph}^{\rm broken}}{T^3 H(T)} = 
\kappa \alpha_W^4 \left(\frac{ 4\pi k}{g_W} \right)^7 e^{- 1.52 k \frac{4\pi}{g_W}} \sqrt{\frac{90}{\pi^2 g_*}} 
\frac{M_P}{T} \geq 1,
 }
leading to the constraint
\dis{
C_2 C_3 \sin\delta_c + C_1 S_2 S_3 \sin (\delta_c +
\dell) \simeq  2.4 \times 10^{-2},
 }
where $\delta_c$ is a Majorana neutrino phase. So, there is an enough parameter space to allow an acceptable $\Delta L$.
 
\section{Conclusion}

Here, my talk on Type-II leptogenesis is centered on: 
\begin{itemize}
\item[(1)] A short discussion on weak CP,  
\item[(2)] Introduction of a new leptogenesis mechanism in theories with SU(2)$\times$U(1) breaking at high temperature,  

\item[(3)] Relation of $\dell$  to the leptogenesis phase in certain CP violation models, 
\item[(4)] Need of one light intermediate scale Majorana neutrino $N_0$ and another neutrino ${\cal N}$ toward the desired lepton number asymmetry $\epsilon_L$.
\item[(5)] Inert Higgs and the SM Higgs mix to provide $\Delta L\ne 0$ vertex in the loop diagram.

\end{itemize}

\acknowledgments{This work is supported in part by the National Research Foundation (NRF) grant funded by the Korean Government (MEST) (NRF-2015R1D1A1A01058449) and  the IBS (IBS-R017-D1-2016-a00).}



\end{document}